\documentclass[twocolumn,preprintnumbers,amsmath,amssymb]{revtex4}

\usepackage{graphicx}
\usepackage{dcolumn}
\usepackage{bm}

\begin{document}



\title{Electrical excitation  of shock and soliton-like   waves
in two-dimensional electron
channels
} 
\author{E.~Vostrikova and A.~Ivanov}
\affiliation{Department of General Physics, Moscow Institute 
of Physics and Technology,
Dolgoprudny 141700,  Russia}
\author{I.~Semenikhin and V.~Ryzhii} 
\email{v-ryzhii@u-aizu.ac.jp}
 \affiliation{Computer Solid State Physics Laboratory, University of Aizu,
Aizu-Wakamatsu 965-8580, Japan}

\date{\today}

\begin{abstract}
We study  electrical excitation of nonlinear plasma waves in heterostructures
with two-dimensional 
electron channels and with split gates, and the propagation of these waves using
hydrodynamic equations for  electron transport coupled
with two-dimensional Poisson equation for  self-consistent
electric potential. The term related to 
electron collisions with impurities
and phonons as well as the term associated with  viscosity
are included into the hydrodynamic equations.
We demonstrate the formation of shock and soliton-like waves
as a result of the evolution of strongly nonuniform initial
electron density distribution. It is shown that the shock wave front
and the shape of soliton-like pulses
pronouncedly depend on the coefficient of viscosity,
the  thickness of the gate layer and the nonuniformity
of the donor distribution along the channel.
The electron collisions result in damping of the shock and soliton-like waves,
while they do not  markedly affect the thickness of
the shock wave front.\\
PACS numbers: 73.21.-b, 73.40.-c, 73.43.Lp, 73.43.Cd
\end{abstract}


\maketitle

\section{Introduction}

Plasma waves, i.e., self-consistent spatiotemporal
variations of the electron density and electric
field in two-dimensional electron gas (2DEG) channels~\cite{1,2,3,4,5} 
can be  used
in different semiconductor heterostructure
devices. One of the most important advantages of
2DEG systems in comparison with 3DEG systems is the possibility
to realize  situations when the characteristic plasma frequency
markedly exceeds the frequency of electron collisions with impurities
and phonons. This is achieved using the selective doping
when  donors are spatially separated from the 2DEG channel,
so that the electron density can be rather high, while the
electron-impurity interaction is weakened.
In  2DEG systems with highly conducting electrodes (gates)
similar to field-effect transistor structures,
the electron density can be effectively controlled (in particular, significantly
increased) by the applied voltage. Another attractive feature 
of the 2DEG system is that at realistic parameters the characteristic
plasma frequency falls in the terahertz  range. This opens up
the prospects of creating  novel terahertz devices, for example, sources
and detectors of terahertz
 radiation, frequency multipliers, and so on~\cite{6,7,8}.
In recent experiments~\cite{9,10,11,12,13,14,15,16},
detection of terahertz radiation in and terahertz
 emission from transistor-like
2DEG systems associated with the resonant excitation
of plasma waves were realized (see also~\cite{17,18,19,20,21}). 
Results of theoretical studies of plasma phenomena in
2DEG systems have been reported in numerous publications.
However, nonlinear plasma
phenomena in these systems are  studied far less extensively.
In particular, the plasma phenomena associated with hydrodynamic
nonlinearities in 2DEG transport were considered in 
Refs.~\cite{22,23,24,25,26,27,28}.
The effect of nonlinearities associated with
different contact effects were considered theoretically 
in Refs.~\cite{29,30,31}. 
Nevertheless, nonlinear properties of 2DEG systems
can themselves be used in device applications.
This and recent progress in experimental studies of terahertz 
plasma phenomena
stimulate a significant interest in
nonlinear plasma effects in different 2DEG systems.
 
In this paper, we study nonlinear plasma phenomena associated
with the electrical excitation of plasma waves in 2DEG systems
using both analytical treatment and numerical modeling.
We consider structures
 with a 2DEG channel supplied
with  side contacts and a system
of highly conducting electrodes (split gates) which provides an opportunity
to control the 2DEG by applying voltage signals.
The structures under consideration are schematically shown in Fig.~1.
We use the hydrodynamic model for  electron
transport along the channel coupled with the 2D Poisson
equation for  self-consistent electric potential in and around the channel.
In contrast to  previous studies of nonlinear waves in  
2DEG systems~\cite{22,23,24,25,26},
we take into consideration the 2D nature of the potential
distributions. This allows us to follow 
the change in nonlinear waves properties with
the transition
from  2DEG systems with the gates very close to
the channel to those with rather remote gates when
the plasmon spectrum essentially varies. 
The effect of the spacing between 2DEG
and highly conducting gates on the linear plasma
properties we studied in many papers beginning with the paper by Chaplik.~\cite{2}
(see, in particular, 
Refs.~\cite{32,33,34,35} where the role of the gates and contacts on
the spectra of linear plasma oscillation  were considered).

As one might expect and is shown below,
the 2DEG systems with different plasmon spectra exhibit different
behavior for nonlinear plasma waves.
This is akin to the difference in properties of gravity surface
waves in shallow and deep water pools~\cite{6,7,22,36}.
Apart from this, in our numerical modeling we focus on  electron
scattering on impurities and  phonons as well as on the 2DEG
viscosity due to electron-electron collisions.

\begin{figure}
\begin{center}
\includegraphics[width=7.5cm]{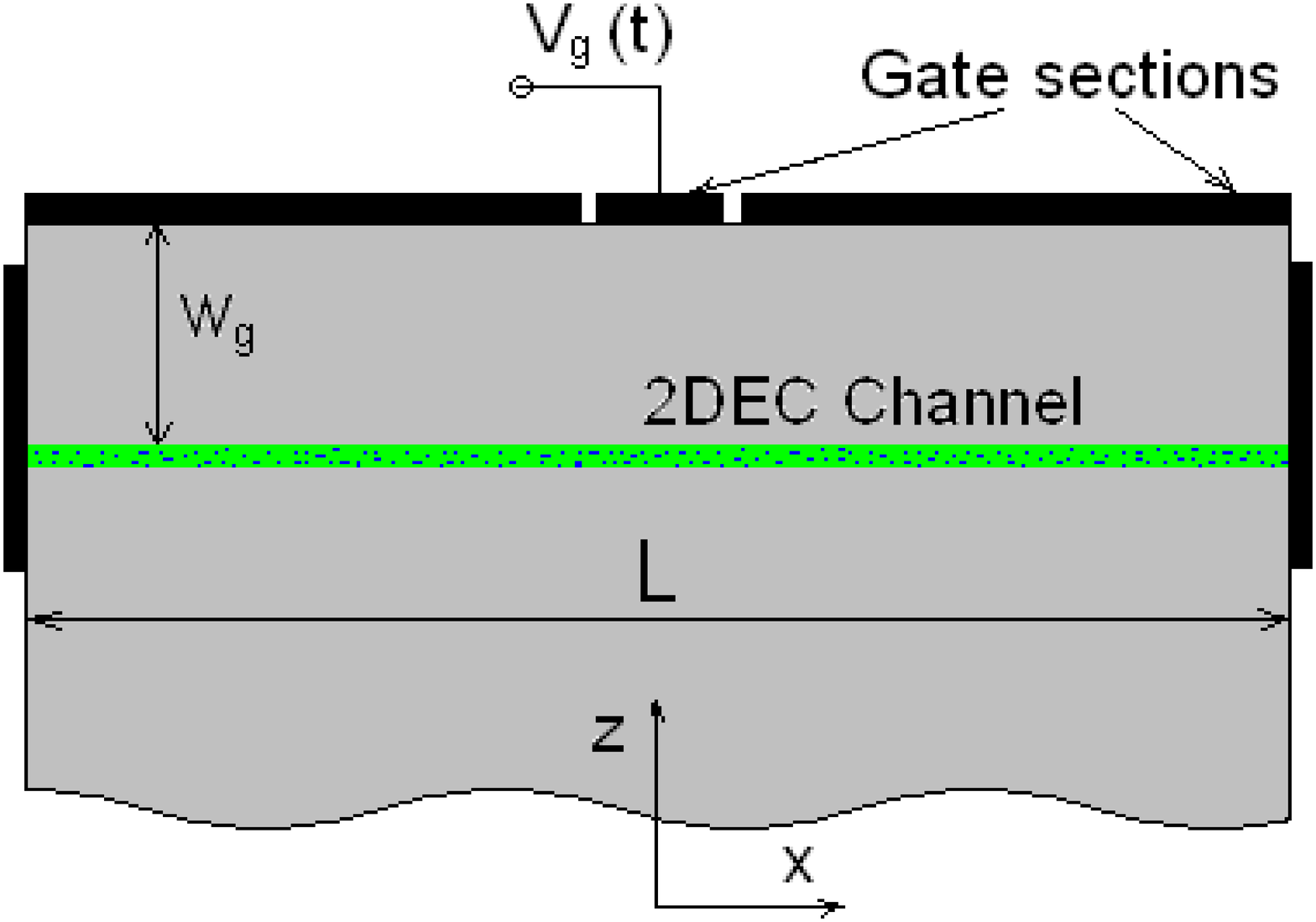}
\includegraphics[width=7.5cm]{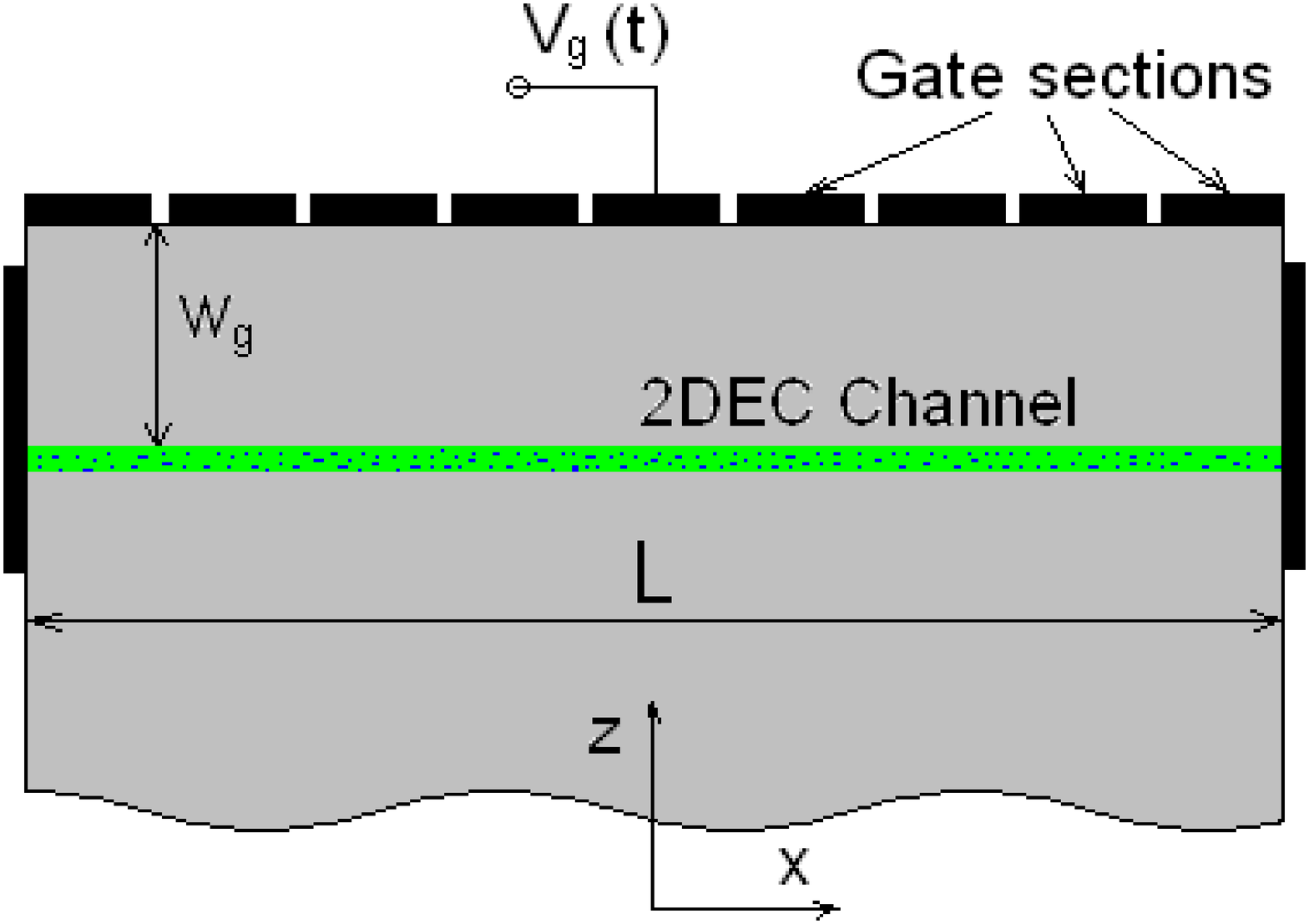}
\end{center}
\label{fig1}
\caption{
Schematic view of heterostructures with 2DEG
channel and with split gate electrode  configurations.
}
\end{figure}

\section{Equations of the model}

We consider the hydrodynamic equations (the Navier-Stokes equation
and the continuity equation) and the Poisson equation
in the following form:

\begin{equation}\label{eq1}
\frac{\partial v}{\partial t}
+ \nu\,v +
v\frac{\partial v}{\partial x} 
- K\frac{\partial^2 v}{\partial x^2}
= \frac{e}{m}\frac{\partial \varphi}{\partial x},
\end{equation}

\begin{equation}\label{eq2}
\frac{\partial \Sigma}{\partial t} +
\frac{\partial \Sigma\,v}{\partial x} = 0,
\end{equation}

\begin{equation}\label{eq3}
\frac{\partial^2 \psi}{\partial x^2}
+ \frac{\partial^2 \psi}{\partial z^2} = 
\frac{4\pi e}{\ae}(\Sigma - \Sigma_d)\cdot\delta(z).
\end{equation}
Here, $v = v(t,x)$ and $\Sigma = \Sigma(t,x)$
are the electron hydrodynamic velocity along  and sheet density in the 2DEG channel,
$\psi = \psi (t,x,z)$ is the electric potential,
$\varphi = \psi (t,x,0)$ is the electric potential in the channel ($z = 0$),
$\Sigma_d = \Sigma_d(x)$ 
is the donor sheet concentration in the 2DEG channel (or near it),
$\nu$ is the frequency of the electron scattering on impurities
and phonons, $K$ is the coefficient of viscosity associated with
the electron-electron scattering processes, 
$e = |e|$ and $m$ are the electron charge and effective mass, respectively,
$\ae$ is the dielectric constant, and
$\delta(z)$ is the Dirac delta function describing the electron confinement
in a relatively narrow channel (its width is assumed to be much smaller
than any lateral size and the distance between the 2DEG channel and
the gate electrode). The axis $x$ is directed along the channel, so that
the coordinates of the side contacts are $x = \pm L/2 \simeq \pm L_g/2$,
where $L$ and $L_g$ are the lengths of the channel and gate (see Fig.~1),
respectively ($L_g \lesssim L$). 
The axis $z$
is directed perpendicular to the 2DEG channel. The latter corresponds to $z = 0$,
whereas the gate electrode corresponds to $z = W_g$, where $W_g$
is the thickness of the gate layer.
In Eq.~(1), we disregard the term corresponding to the gradient 
of the pressure in 2DEG in comparison with the electric force 
(see, for instance,
Ref.~\cite{25}).
The boundary conditions follow from the assumption that
the electron sheet densities at the side contacts are fixed and
that
the electric potentials at the electrodes (the side contacts to the channel and
the gate sections) are  given functions of time.
For the electric potential at nonconducting surfaces (between the electrodes)
linear approximation is used. It is also assumed that
at $z \rightarrow - \infty$, the electric field 
$\partial \psi/\partial z \rightarrow 0$.
The equations of the model under consideration 
account for
the generally  two-dimensional character
of the potential spatial distribution, as well as the electron
transport in the 2DEG channel considering  electron scattering and
 viscosity.

In the case of the structures with relatively small thickness of
the gate layer $W_g$ ($W_g \ll \lambda$, where $\lambda$ is the characteristic
length of the electron density and electric potential nonuniformities),
the Poisson equation (3) can be replaced by the following:

\begin{equation}\label{eq4}
\frac{\varphi - V_g}{W_g}
= 
\frac{4\pi e}{\ae}(\Sigma_d - \Sigma).
\end{equation}

We shall consider the following case.
It is assumed that 
generally the bias dc voltage $V_{g}$ is   applied 
to the  side gate sections.
Apart from this,
a relatively high transient voltage  $V_g(t)$
is applied to
the central
section, so that $V_{g}(t) = V_{g} + \Delta V_g[\Theta(-t) +
\Theta(t)\exp (-t/t_0)]$,
where $\Theta(t)$ is the unity step function
and $t_0$ is the characteristic time of the gate section recharging.
The recharging time $t_0$
is determined
by the capacitance, $C_0$, of the gate section and the pertinent resistance, $R_0$,
of the control circuit: $t_0 = R_0C_0$.
The bias gate voltages  
result in a nonuniform 
distribution of the electron density along the channel,
which at $t < 0$
exhibits a
maximum in the channel center. We shall  mainly consider
2DEG systems with uniformly distributed  donors in the channel 
(or in the gate layer slightly above the channel).
The case of strongly nonuniform doping along the channel
will be briefly studied as well (in Sec.~V).

The linearized version of Eqs.~(1) - (3) governs the propagation of
the plasma waves with the dispersion  relation, which in
the case $V_g = const$ is
given by the following formula: 
\begin{equation}\label{eq5}
\omega[\omega + i(\nu + Kq^2)] = \frac{s_0^2\,q}{W_g[\coth(|q|W_g) + 1]}.
\end{equation}
Here, $\omega$ and $q$ are the frequency and wavenumber of a linear plasma wave,

\begin{equation}\label{eq6}
s_0 = \sqrt{
\frac{4\pi e^2\Sigma_0W_g}{\ae m}}
\end{equation}
is the characteristic velocity of the plasma wave in the gated 2DEG channel,
and 
\begin{equation}\label{eq7}
\Sigma_0 = \Sigma_d + \frac{\ae V_g}{4\pi eW_g}
\end{equation}
is the ac electron sheet density.
In the limits of long waves ($qW_g \ll 1$) and short waves ($qW_g \gg 1$),
Eq.~(5) yields the following relationships~\cite{1,3}
\begin{equation}\label{eq8}
{\rm Re}\,\omega \simeq s_0\,q\biggl(1 - \frac{|q|W_g}{2}\biggr)
\end{equation}
and
\begin{equation}\label{eq9}
{\rm Re}\,\omega \simeq s_0\,\sqrt{\frac{|q|}{2W_g}}.
\end{equation}
The dispersion relations given by Eqs.~(5), (8), and (9)
are similar to those for gravity waves on the surface
of a water channel. In particular, Eq.~(8) corresponds to
the so-called ``shallow water'' case, whereas Eq.~(9) corresponds
to the ``deep water'' case (see, for instance,~\cite{6,7,22,36}).
This is a consequence of the analogy of Eqs.~(1) - (3) and
the pertinent hydrodynamic equations for waves on the surface
of a liquid under
gravity force. The  damping rate of the plasma wave
(in the case $\nu + Kq^2 \ll {\rm Re}\,\omega$)
is given by
\begin{equation}\label{eq10}
{\rm Im}\,\omega \simeq - \frac{1}{2}(\nu + Kq^2).
\end{equation}

\section{Riemann solution and shock wave formation}

Consider first Eqs.~(1), (2), and (4)
assuming  that the gate layer is sufficiently thin,
the length of the channel is sufficiently large, so that we
may disregard
the boundary conditions,
and  that the electron collisions with
impurities and phonons as well as the viscosity
can be neglected.
In this case, we arrive at the following equations:
\begin{equation}\label{eq11}
\frac{\partial v}{\partial t} +
v\frac{\partial v}{\partial x} 
= \frac{e}{m}\frac{\partial \varphi}{\partial x},
\end{equation}

\begin{equation}\label{eq12}
\frac{\partial \Sigma}{\partial t} +
\Sigma\,\frac{\partial v}{\partial x} = - v\,\frac{\partial \Sigma}{\partial x},
\end{equation}
\begin{equation}\label{eq13}
\frac{e\varphi}{m}
= 
s_0^2\biggl(1 - \frac{\Sigma}{\Sigma_0}\biggr).
\end{equation}
The system of Eqs.~(11) - (13) can be presented as

\begin{equation}\label{eq14}
\frac{\partial v}{\partial t} +
v\frac{\partial v}{\partial x} 
= - \frac{s_0^2}{\Sigma_0}\frac{\partial \Sigma}{\partial x},
\end{equation}

\begin{equation}\label{eq15}
\frac{\partial \Sigma}{\partial t} +
\Sigma\,\frac{\partial v}{\partial x} = - v\,\frac{\partial \Sigma}{\partial x}.
\end{equation}
As pointed out above, Eqs.~(14) and (15) are akin to
the equations governing gravity waves in shallow water channels.
They are also identical in form to the equations governing
isentropic flows of a compressible gas with the adiabatic index $\gamma = 2$.
This indicates that Eqs.~(14) and (15) have solutions
in the form of the Riemann waves which, under certain conditions, can
transform into  shock waves~\cite{36,37}.
One needs also to point out that the solution of Eqs.~(14) and (15)
should describe nonlinear waves with the front steepening 
with time until the moment when the solution becomes three-valued. 
Analyzing Eqs.~(14) and (15), we shall use a standard approach
(see, for instance,~\cite{37,38}), assuming that $v = v(\Sigma)$.
Considering the latter, Eqs.~(14) and (15) become as follows:
 \begin{equation}\label{eq16}
\frac{d\,v(\Sigma) }{d\,\Sigma}\biggl[\frac{\partial \Sigma}{\partial t}
+ v(\Sigma)\,\frac{\partial \Sigma}{\partial x}\biggr]
= - \frac{s_0^2}{\Sigma_0}\frac{\partial \Sigma}{\partial x},
\end{equation}

\begin{equation}\label{eq17}
\frac{\partial \Sigma}{\partial t}
+ v(\Sigma)\,\frac{\partial \Sigma}{\partial x} = 
- \Sigma\frac{d\,v(\Sigma)}{d\,\Sigma}\frac{\partial \Sigma}{\partial x}.
\end{equation}
Equations (16) and (17) are identical if
\begin{equation}\label{eq18}
\Sigma \biggl[\frac{d\,v(\Sigma)}{d\,\Sigma}\biggr]^2
= \frac{s_0^2}{\Sigma_0}.
\end{equation}
Equation~(18) results in 
\begin{equation}\label{eq19}
\frac{d\,v(\Sigma)}{d\,\Sigma}
= \pm \frac{s_0}{\sqrt{\Sigma_0\Sigma}}.
\end{equation}
Considering the solution corresponding to the sign ``+'' 
in Eq.~(19), and introducing 
\begin{equation}\label{eq20}
s = s_0\sqrt{\frac{\Sigma}{\Sigma_0}},
\end{equation}
we obtain

\begin{equation}\label{eq21}
v(\Sigma) = 2s + const.
\end{equation}
Assuming that $\Sigma = \Sigma_0$ at $v = 0$,
we find 
\begin{equation}\label{eq22}
v(\Sigma) = 2(s - s_0) = 2s_0\biggl(\sqrt{\frac{\Sigma}{\Sigma_0}} - 1\biggr).
\end{equation}
Using Eq.~(22) with Eqs.~(16) and (19),
we arrive at the Riemann solution:

\begin{equation}\label{eq23}
x  = [v(\Sigma) + s(\Sigma)]t + C(\Sigma),
\end{equation}
where $s(\Sigma)$ is given by Eq.~(20), and $C(\Sigma)$
is determined by the initial conditions. Hence,
the electron density as a function of $x$ and $t$
is given (in implicit form) by the following equation
\begin{equation}\label{eq24}
x  = s_0\biggl(3\sqrt{\frac{\Sigma}{\Sigma_0}} - 2\biggr)t + C(\Sigma).
\end{equation}
Using Eq.~(24),
we can find the wave breaking time $t_{br}$, i.e.,
the time of formation of the discontinuity or the time
of formation of a shock wave. This time is determined by
the following conditions~\cite{36}: 
\begin{equation}\label{eq25}
\frac{\partial x}{\partial \Sigma}\biggr|_{t = t_{br}} = 0, \qquad
\frac{\partial^2 x}{\partial \Sigma^2}\biggr|_{t = t_{br}} = 0.
\end{equation}
Equations~(25) yield
\begin{equation}\label{eq26}
t_{br} = - \frac{2}{3} \frac{\sqrt{\Sigma_0\Sigma}}{s_0}
\frac{dC(\Sigma)}{d\Sigma}\biggr|_{t = t_{br}}
\end{equation}
and
\begin{equation}\label{eq27}
t_{br} = \frac{4}{3} \frac{\Sigma_0^{1/2}\Sigma^{3/2}}{s_0}
\frac{d^2C(\Sigma)}{d\Sigma^2}\biggr|_{t = t_{br}}.
\end{equation}

As an example, let us consider
the case when the initial distribution of the electron
density is as follows:
\begin{equation}\label{eq28}
\Sigma|_{t = 0} = \Sigma_0 + 
\Delta\Sigma_0\exp \biggl[- \biggl(\frac{x}{a}\biggr)^2\biggr],
\end{equation}
where
$\Delta \Sigma_0$ is the magnitude of the electron density
perturbation and $a$ is its characteristic length.
This yields
\begin{equation}\label{eq29}
C(\Sigma) = a\sqrt{\ln\biggl(\frac{\Delta\Sigma_0}
{\Sigma - \Sigma_0}\biggr)}.
\end{equation}
According to Eq.~(22), this distribution
of the electron density corresponds to the following
initial velocity distribution: 
\begin{equation}\label{eq30}
v|_{t = 0} = 2s_0\biggl\{\sqrt{1 + 
\frac{\Delta\Sigma_0}{\Sigma_0}\exp \biggl[- \biggl(\frac{x}{a}\biggr)^2\biggr]}
 - 1\biggr\}.
\end{equation}
If
\begin{equation}\label{eq31}
\Sigma|_{t = 0} = \Sigma_0 - 
\frac{2\Delta\Sigma_0}{\pi}\tan^{-1}\biggl(\frac{x}{a}\biggr),
\end{equation}
one obtains
\begin{equation}\label{eq32}
C(\Sigma) = a\,\tan\biggl[\frac{\pi}{2}
\biggl(\frac{\Sigma_0 - \Sigma}{\Delta\Sigma_0}\biggr)\biggr].
\end{equation}

Figures~2 and 3  show the transformation of the Riemann waves governed
by Eq.~(23) with the initial conditions given  by Eqs.~(28) and
(31), respectively, up to the moments  when conditions~(26) and (27)
are satisfied (up to $t = t_{br}$). These moments correspond 
to the rightmost  curves in Figs.~2 and 3.
The electron density and  the coordinate in Figs.~2 and 3
are normalized by
$\Sigma_0$ and $s_0/\nu$, respectively. The obtained spatiotemporal
variations of $\Sigma = \Sigma(t,x)$ and $v = v(\Sigma(t,x))$
were compared with the results of direct numerical solution
of Eqs.~(14) and (15), and their coincidence was confirmed.   

\begin{figure}
\begin{center}
\includegraphics[width=6.5cm]{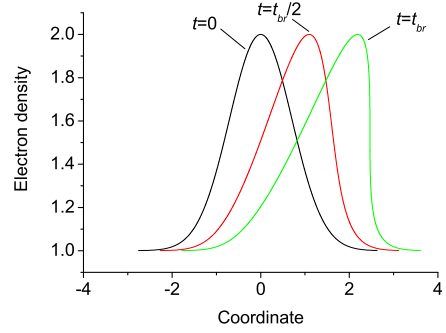}
\end{center}
\label{fig2}
\caption{
Transformation of spatial distribution  o normalized electron density 
$\Sigma/\Sigma_0$
in the Riemann wave governed by Eq.~(23) 
with initial condition Eq.~(28) and $\Delta \Sigma_0/\Sigma_0 = 1$.
}
\end{figure} 

\begin{figure}
\begin{center}
\includegraphics[width=6.5cm]{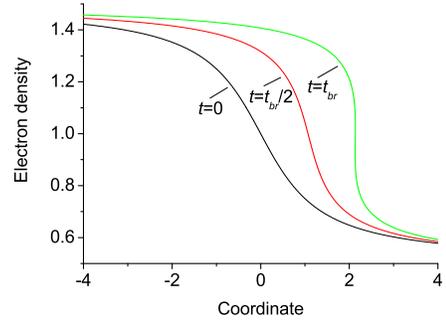}
\end{center}
\label{fig3}
\caption{
The same as in Fig.~2 but
for initial condition given by  Eq.~(31)
and $\Delta \Sigma_0/\Sigma_0 = 0.5$.
}
\end{figure}

The solutions obtained above are related to the formation
of  shock waves with zero front thickness.
The inclusion of  electron collisions and  viscosity
(that correspond to  real systems)
might affect the shape of the shock wave front.
At a sufficiently sharp wave front, i.e., at a small thickness
of the front $\delta$, 
the viscosity term $- Kq^2v$ in Eq.~(1)
 markedly exceeds the term associated with the electron collisions
$\nu\,v$. Indeed, comparing these two terms and 
setting $q \sim \delta^{-1}$,
one can find that the effect of viscosity dominates when
$\delta < \sqrt{K/\nu}$. Using Eqs.~(1), (12), and (13) (valid in the case
of small gate layer thickness) and
following the standard procedure~\cite{36}, 
the shock wave thickness determined by the viscosity
can be estimated as~$\delta \simeq K/s_0$.
Substituting $\delta$ from this equation into the above inequality,
we obtain the condition when the viscosity prevails the collisions
in the formation of the shock wave front: $K \ll s_0^2/\nu = K_c$.
Assuming $s_0 = 10^8$~cm/s and $\nu = 10^{12}$~s$^{-1}$,
we have $K_c \sim 10^4$~cm$^2$/s. Since the characteristic
value of the coefficient of viscosity in real 2DEG systems
is  $K \sim \hbar/m$ (see, for instance, Refs.~\cite{6,7}), 
where $\hbar$ is the reduced Planck constant, for the GaAs channel
one obtains
$K \sim 15$~cm$^2$/s. This implies that $K \ll K_c$
and that the electron collisions
should  not affect the shape of the shock wave front
as confirmed by   the results on numerical
modeling in the following.

In  2DEG systems where the gate layer
is not too
thin,
the thickness of the wave front $\delta$ can be determined  by
the deviation of the plasma wave dispersion relation from the linear one (i.e.,
by 2D effects leading to a difference in the group
velocities of the harmonics with different
wavenumber $q$) rather than by the viscosity. 
Comparing the contributions of these two mechanisms, 
one can find that 
the first mechanism dominates if 
$K < s_0W_g = K_d$ with $K_d \simeq K_c(W_g\nu/s_0)$.

Thus, in  2DEG systems with moderate gate layer thickness,
the nonlinear waves should be fairly smooth, so that
a soliton-like wave can form~\cite{27,28}  rather than a shock wave. 
This is confirmed by the results of
 numerical simulations based on the general
system of Eqs. ~(1) - (3) demonstrated in the next section.

\section{Shock and soliton-like waves: Numerical modeling}

Since an analysis of nonlinear regimes governed
by
 the general equations of our model,
which accounts for both
the electron collisions and viscosity,  
i.e. Eqs.~(1) - (3), can not be realized analytically,
we solved these equations numerically considering for definiteness
the upper structure in Fig.~1.
The following  dimensionless variables are used:
time $\tau = t\nu$, coordinates $\xi = x\nu/s_0$ and  
$\eta = z\nu/s_0$,
velocity
$u = v/s_0$, electron density
$\sigma = \Sigma/\Sigma_0$, and electric potential $\phi = e\varphi/ms_0^2$.
In these variables, the system of Eqs.~(1) - (3) is characterized by the following
parameters: ${\cal L}_g = L_g\nu/s_0$ and
$w_g = W_g\nu/s_0 = (W_g/L_g){\cal L}_g$,
$\tau_0 = t_0\,\nu$, and $k = K\,\nu/s_0^2$.  
The Galerkin spectral method~\cite{39} was used in 
numerical calculations.

Figures~4 - 7 demonstrate the evolution
of the spatial distributions of the electron density (normalized by
$\Sigma_0$) 
in the structures with ${\cal L}_g = 10$ and
different normalized gate layer thicknesses 
($w_g = 0.001 - 1$) 
in response to a drastic change of the central gate section
potential: $\Delta v_g = 2\exp (- 100\tau)$ ($\tau_0 = 10^{-2}$).
It is assumed that $k = 2\times 10^{-3}$. These cases correspond
to the spatial distributions of the electron density
at the initial moment with a maximum at $x = 0$, where 
$(\Sigma/\Sigma_d)|_{t = 0} \simeq 1.3 - 3$.
If $s_0 = 10^8$~cm/s and $\nu = 10^{12}$~s$^{-1}$,
the results demonstrated in Figs.~4 - 7 correspond to
the gate length $L_g = 10~\mu$m 
(i.e., to a fairly long channel) and 
$W_g/L_g = 10^{-4} - 10^{-1}$. The length of the central gate section
is about $0.5~\mu$m.
As shown, in  systems with small values of the  parameter $w_g$ (Figs.~4 and 5),
the initial perturbation of the electron density
transforms into two pulses propagating from the channel center
to the side contacts. After a short time, 
the front of these pulses becomes
sharp, so  the pulses turn into shock waves.
The results obtained for the case of small $w_g$
coincide with the results of the calculations by Dmitriev
{\it at al.}~\cite{23}
and  Rudin {\it at al.}~\cite{24,25}.
However, in 2DEG  systems with larger values of the parameter $w_g$,
the propagating waves have the form of relatively smooth pulses 
with a soliton-like
shape (Fig.~6). 
At sufficiently large $w_g$, the transformation
of the initial perturbation of the electron density (Fig.~7) 
is  even qualitatively different from that
in the case of effectively gated 2DEG systems.
This difference in  the nonlinear waves propagated
in 2DEG systems with different parameter $w_g$ can be attributed to
a significant difference in
the dispersion relations of the plasma waves in the cases
of small and large spacings $W_g$  (small and large values of parameter $w_g$).

\begin{figure}
\begin{center}
\includegraphics[width=8.0cm]{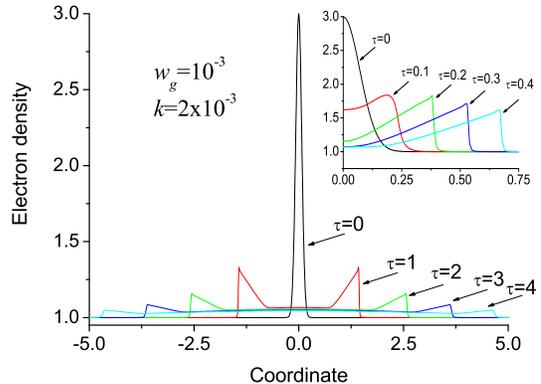}
\end{center}
\label{fig4}
\caption{
Evolution of spatial distributions of normalized electron density
$\Sigma/\Sigma_0$
in the 2DEG system with normalized
gate layer thickness
$w_g = 10^{-3}$.
The inset corresponds to the early stage when the shock
wave is formed. 
This case is analogous to the propagation of surface waves
in a ``shallow water'' channel. 
}
\end{figure} 

\begin{figure}
\begin{center}
\includegraphics[width=8.0cm]{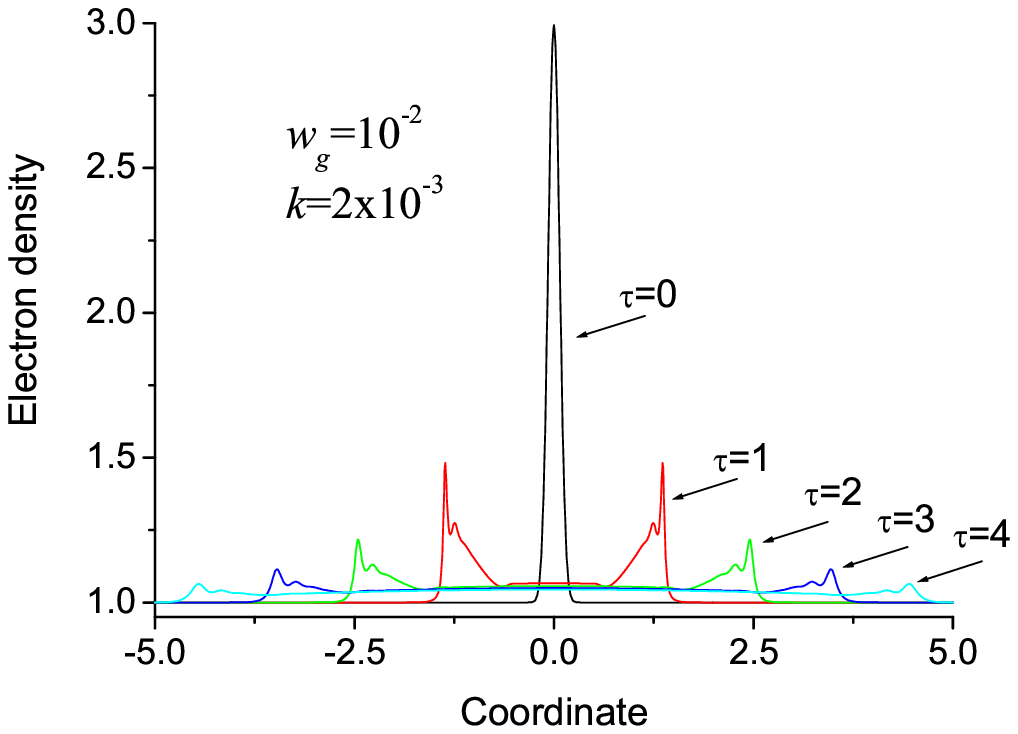}
\end{center}
\label{fig5}
\caption{
The same as in Fig.~4 but for $w_g = 10^{-2}$
}
\end{figure} 

\begin{figure}
\begin{center}
\includegraphics[width=8.0cm]{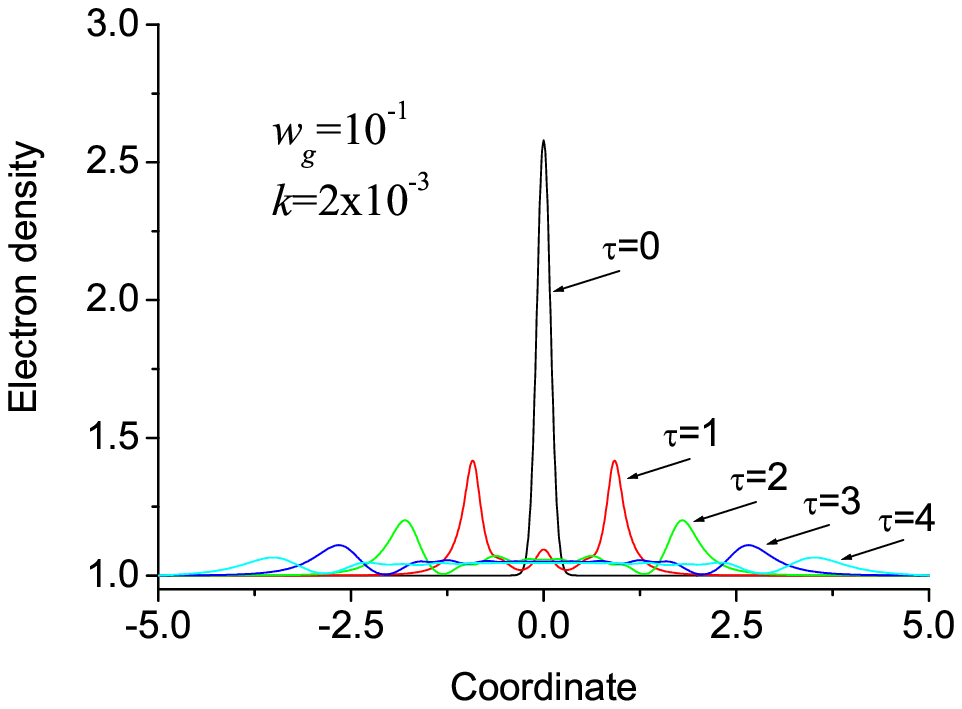}
\end{center}
\label{fig6}
\caption{
The same as in Fig.~4 but for $w_g = 10^{-1}$.
}
\end{figure} 

\begin{figure}
\begin{center}
\includegraphics[width=8.0cm]{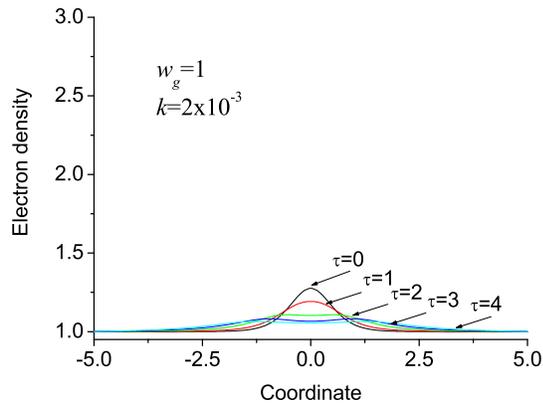}
\end{center}
\label{fig7}
\caption{
The same as in Fig.~4 but for $w_g = 1$ (thick gate layer - ``deep
water'' case).
}
\end{figure}

As seen from Figs.~4 - 7, the amplitude of the propagating pulses 
markedly decreases with time. This effect is obviously associated with 
electron collisions.
Comparing Figs.~4, 5,  and 6, one can see that
the thickness of the wave front markedly increases with increasing
 $w_g$ characterizing the role of the plasma wave dispersion
associated with the 2D nature of the distributions of the electrical
potential in the systems under consideration.

Figure~8 shows the evolution of
the electron density distributions in a 2DEG system
with  parameter $k$ substantially smaller than 
in Figs.~4 - 7. One can see that in this case,
the electron density distributions behind the shock wave front
exhibit pronounced oscillations.
To avoid  numerical calculation artifacts,
the pertinent modeling was conducted with rather different
numbers of modes in the Fourier transformation 
(varied from  $2\times10^3$ to $2\times10^4$). 
The independence of the result of the number of the modes 
used in the calculations was
confirmed.

\begin{figure}
\begin{center}
\includegraphics[width=8.0cm]{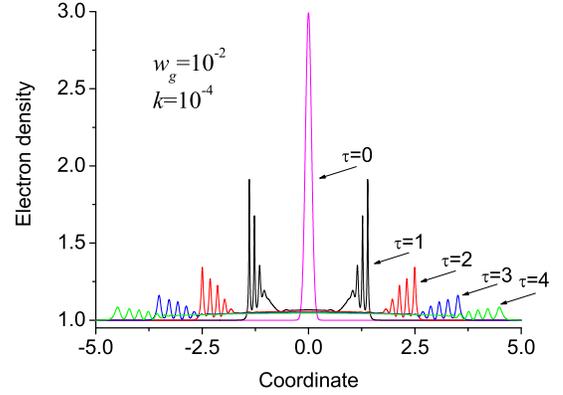}
\end{center}
\label{fig8}
\caption{Snapshots of  spatial distributions of normalized
electron density $\Sigma/\Sigma_0$
with oscillatory structure behind the shock wave front
in the system  with normalized
gate layer thickness
$w_g = 10^{-2}$ and small viscosity ($k = 10^{-4}$).
}
\end{figure}

The oscillatory character of the electron density distributions
 can be attributed
to the instability 
of the shock waves with relatively smooth (but nonuniform)
distributions behind the shock wave front.
As shown previously~\cite{40}, the  nonuniformity of 2DEG in the channel
 can markedly
affect the conditions of plasma instability.
Considering the nonuniformity of the electron density distribution
in a propagating shock wave and assuming that the characteristic length
of this perturbation $q^{-1}$ is small in comparison with
the scale of the nonuniformity
of the electron density distribution behind the front,
one can get the following equation for the damping rate of
the electron density perturbations behind the front 
(compare with Eq.~(10)): 
\begin{equation}\label{eq33}
{\rm Im} \,\omega \simeq  
- \frac{1}{2}\biggl(\nu + Kq^2 - s_0\frac{d\ln \Sigma_0}{q\, d x}\biggr).
\end{equation}
\begin{figure}
\begin{center}
\includegraphics[width=8.0cm]{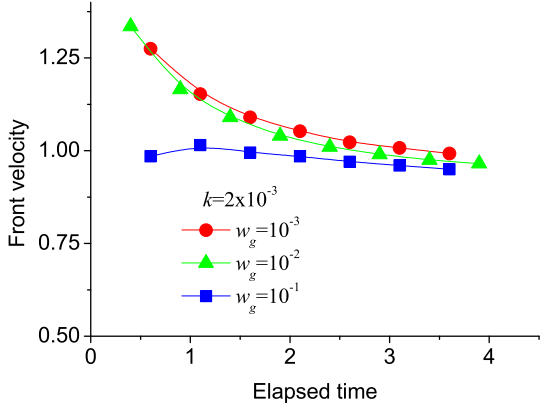}
\end{center}
\label{fig9}
\caption{Time dependences of the front velocity for
the same parameters as in Figs.~4 - 6. 
}
\end{figure}

\begin{figure}
\begin{center}
\includegraphics[width=8.5cm]{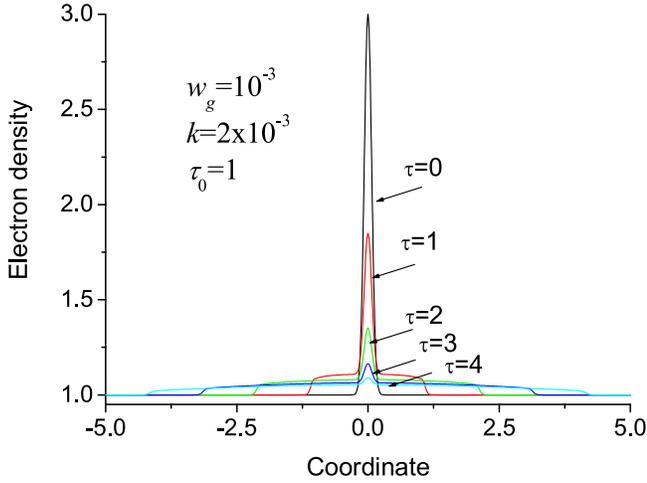}
\end{center}
\label{fig10}
\caption{Evolution of spatial distribution of normalized electron density 
$\Sigma/\Sigma_0$ in  the 2DEG system with the same parameters as in Fig.~4 
but at relatively slow voltage switch-off ($\tau_0 = 1$).
}
\end{figure}
In the case  Im\,$\omega > 0$, i.e., when 
\begin{equation}\label{eq34}
\frac{d\ln \Sigma_0}{d x} > \frac{\nu + Kq^2}{s_0},
\end{equation}
the nonuniform but relatively smooth
electron density 
 distribution behind the
shock wave front can become unstable resulting in the formation
of  oscillatory distributions.
This instability condition shows that
the electron density distribution
can be unstable when the gradient of the electron density is sufficiently
large. 
Since the magnitude of the electron density peak decreases
with  wave propagation (due to  electron collisions),
and the total number of electrons is constant, the electron density
gradient decreases as well. As a result,
inequality~(34) can be satisfied for the perturbations with
the spatial period increasing with time. Just such a behavior
of the oscillatory structure behind the shock wave front is seen in Fig.~8.

Figure~9 shows how the velocity of the wave front (normalized by $s_0$)
changes with time (normalized by $\nu^{-1}$). We define the 
velocity of the wave front  as
the velocity of movement of the point where 
$\Sigma = (\Sigma_{max} + \Sigma_0)/2$.
Here,  $\Sigma_{max}$ is the peak value of the electron density.
One can see that in  2DEG systems with relatively thin gate layers
in which the formation of the shock waves occurs
(corresponding to Figs.~4 and 5), the velocity of the front movement
exceeds the characteristic plasma wave velocity $s_0$
but this front movement decelerates with elapsed time.
This can be attributed to 
the fact that, according to
the general properties
of shock waves~\cite{35} (see also Ref.~\cite{25}), their velocity
is determined by $s_0$ and $s = s_0\sqrt{\Sigma/\Sigma_0}$
(see Eq.~(20)) with the latter value decreasing throughout
the wave propagation (as seen in Figs.~4 and 5).
However, the velocity of propagation of more smooth pulses (see curve marked
 by squares in Fig.~9)
 in   the case of larger gate thickness 
 is close to $s_0$.

The results shown in Figs.~4 - 8 correspond to relatively fast
switch-off of the voltage applied to the central gate section
($\tau_0 = 10^{-2}$). The obtained results do not change
when $\tau_0$ increases up to $\tau_0 = 10^{-1}$.
However, at relatively slow voltage switch-off, i.e. at moderate values
of parameter $\tau_0$,
the pattern of  relaxation of  electron density
becomes markedly different. Figure~10 shows
the relaxation  of  electron density 
in  the 2DEG system with the same parameters as in Fig.~4 
at $\tau_0 = 1$.
In this case, the transformation of the electron density distribution
 closely resembles
  the density peak smearing considered previously in the framework
of the conductive model~\cite{22,41}.

\section{Wave propagation in nonuniform channel: Effect of ``tsunami''}
\begin{figure}
\begin{center}
\includegraphics[width=8.5cm]{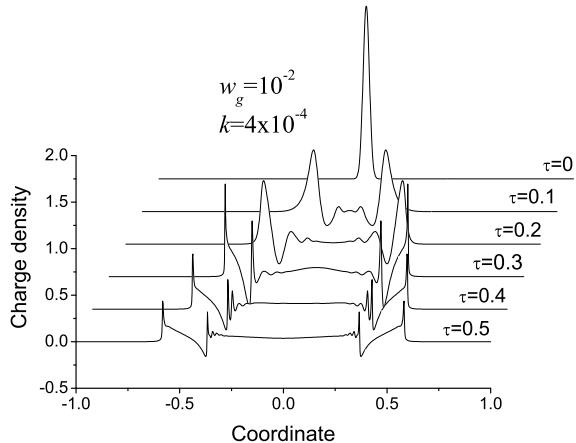}
\end{center}
\label{fig11}
\caption{Evolution of spatial distribution of normalized
sheet charge $(\Sigma - \Sigma_0)/\langle \Sigma_0 \rangle$
in the 2DEG system with nonuniform donor distribution.
}
\end{figure}

Consider now a 2DEG system with markedly nonuniform distribution
of donors $\Sigma_d = \Sigma_d(x)$ along the channel, so that the quantity
$\Sigma_0$ given by Eq.~(7) also depends on the coordinate $x$. We assume
that 

\[\Sigma_d = \left\{\begin{array}{ll}
\Sigma_d^{c} & \mbox{if $|x| < L_0$,}\\
\Sigma_d^{e} + (\Sigma_d^{c} - 
\Sigma_d^{e})e^{- \biggl(\displaystyle\frac{2|x| - L_d}{2a_d}\biggr)^2} &
\mbox{if $|x| > L_0$,}
\end{array} \right.\]
where $\Sigma_d^{c}$ is the sheet density of donors in the channel
center,
$\Sigma_d^{e}$ is the donor density near its edges, 
and $L_d$ and $a_d$
are the characteristic lengths of the spatial distribution of donors
 ($L_d < L$).
Figure~10 shows the evolution of the spatial distribution  of the
sheet charge density in the 2DEG channel normalized by 
$\langle \Sigma_0 \rangle$, where the symbol  $\langle...\rangle$
means the averaging over the channel length.
One can see that two 
soliton-like waves propagating in opposite directions
toward the side contacts
arise. 
However, upon reaching the regions with relatively low
donor density ($\tau \sim 0.25$), 
these waves transform into shock waves with rather sharp
fronts and complex structures behind the front.
In particular, the fronts are followed by depletion regions
and then by the secondary shock waves.
The  behavior pattern  of the wave originated from
the initial pulse resembles  tsunami 
in the ocean. This is due to an analogy between the equations
governing the 2DEG in the systems under consideration
and the equations describing the  gravity surface
waves in shallow and deep water pools.

\section{Conclusions}
We studied   electrical excitation 
of nonlinear plasma waves in heterostructures
with 2DEG channels and split gates.
The propagation of these waves was considered
both analytically and using numerical modeling.
The hydrodynamic electron transport model accounted for
electron collisions with impurities and phonons
as well as  2DEG viscosity.
The hydrodynamic equations were supplemented by
the 2D Poisson equation for the self-consistent electric
potential.
We demonstrated the formation of shock and soliton-like waves
as a result of the evolution of strongly nonuniform initial
electron density distribution. It was found that the shock wave front
and the shape of soliton-like pulses
pronouncedly depend on the coefficient of 2DEG viscosity,
the  thickness of the gate layer, and the nonuniformity
of the donor distribution along the channel.
In the case of the 2DEG channel with strongly nonuniform doping,
the effect of transformation of relatively smooth waves into
shock waves, resembling a tsunami effect, was observed. 
The electron collisions result in damping of the shock 
and soliton-like waves.
However, they do not markedly affect the thickness of
the shock wave front.
Due to the fairly sharp wavefront of  nonlinear waves in the 2DEG
systems under consideration or due to their strongly oscillatory structure
(as shown in Fig.~8),
the transient charges induced by 
the propagating charges of these waves  in the side contacts and, hence,
in the external circuit  might exhibit steep surges.
Fast transient currents
in such a circuit including an antenna might provide
generation of electromagnetic
radiation with relatively high
frequencies.The heterostructures with a nonuniform doping of 2DEG
channel appears to be rather promising.
In the structures with the gate separated into several short
sections (see the lower structure in Fig.~1) 
the propagating charges of shock waves might
result in even steeper surges of the charges induced
in these section~\cite{42}. Since these sections can serve as
an antenna components, fast variations of the induced
charges and,
hence, the dipole momentum, can lead to emission
of electromagnetic radiation. There are several characteristic
frequencies related to the situation:
$f_1 \sim s_0/L_g$, $f_2 \sim s_0/l_g$,
where $l_g$ is the lateral spacing between the neighboring
gate sections, and $f_3 \sim s_0/\delta$.
The frequencies $f_1$, $f_2$, and $f_3$ can be much larger
that the inverse recharhing time $t_0^{-1}$
easily fall
 in the terahertz range at realistic values of the 
characteristics lengths due to fairly large values
of the characteristic velocity
of plasma wave $s_0$ (about $10^8$~cm/s).
The mechanism of the generation
of relatively high frequency electrical signals,
which convert into electromagnetic radiation is akin to
the frequency multiplication due to the plasma nonlinearity. 
Thus, the excitation of shock and soliton-like waves
in the heterostructures under consideration by external electrical
signals  might be used to generate terahertz radiation.                  

\section*{Acknowledgments}
The authors are grateful to R.~Gupta and A.~Satou for comments on
the manuscript.
One of the authors (V.R.) thanks T.~Otsuji, A.~Chaplik, and
M.~S.~Shur for stimulating discussions.
This work was supported by the  Grant-in-Aid for Scientific Research (S)
from the Japan Society for Promotion of Science, Japan.



\end{document}